\documentclass[reqno]{amsart}

\usepackage{amsmath,amsfonts}
\usepackage{epsfig}
\usepackage{graphicx}
 \usepackage{verbatim} 

\def\alphaset{{\mathfrak A}}
\def\Av{{\rm Av}}

\def\En{{\rm En}}
\def\Gspace{{\mathfrak G}}

\def\opDelta{\widehat{\Delta}}
\def\opB{\widehat{B}}

\def\Tr{{\rm Tr}}

\def\tr{{\rm Tr}}

\def\N{{\mathbb N}}

\def\R{{\mathbb R}}

\def\uu{{\underline{u}}}

\def\ux{{\underline{x}}}

\def\cH{{\mathcal H}}

\def\cL{{\mathcal L}}
\def\cN{{\mathcal N}}

\def\1{{\bf 1}}

\def\eqnn{\begin{eqnarray*}}
\def\eeqnn{\end{eqnarray*}}
\def\eqn{\begin{eqnarray}}
\def\eeqn{\end{eqnarray}}

\def\prf{\begin{proof}}
\def\endprf{\end{proof}}

\theoremstyle{plain}
\newtheorem{theorem}{Theorem}[section]
\newtheorem{definition}[theorem]{Definition}

\newtheorem{lemma}[theorem]{Lemma}

\newtheorem{remark}[theorem]{Remark}

\numberwithin{equation}{section}

\begin{document}

\parskip=8pt

\title[Blowup solutions for the GP hierarchy]
{Energy conservation and blowup of solutions for focusing Gross-Pitaevskii hierarchies}

\author[T. Chen]{Thomas Chen}
\address{T. Chen,  
Department of Mathematics, University of Texas at Austin.}
\email{tc@math.utexas.edu}

\author[N. Pavlovi\'{c}]{Nata\v{s}a Pavlovi\'{c}}
\address{N. Pavlovi\'{c},  
Department of Mathematics, University of Texas at Austin.}
\email{natasa@math.utexas.edu}

\author[N. Tzirakis]{Nikolaos Tzirakis}
\address{N. Tzirakis,  
Department of Mathematics, University of Illinois at Urbana-Champaign.}
\email{tzirakis@math.uiuc.edu}


\begin{abstract}
We consider solutions of the focusing cubic and quintic Gross-Pitaevskii (GP) hierarchies.
We identify an observable corresponding to the 
average energy per particle, and we prove that it is a conserved quantity. 
We prove that all solutions to the focusing GP hierarchy at the $L^2$-critical 
or $L^2$-supercritical level blow up in finite time
if the energy per particle in the initial condition is negative. 
Our results do not assume any factorization of the initial data.
\end{abstract}

\maketitle

\section{Introduction}

The mathematical analysis of interacting Bose gases is a rich
and fascinating research topic that is experiencing remarkable progress in recent years. 
One of the fundamental questions in this field
concerns the mathematically
rigorous proof of Bose-Einstein condensation; 
for some recent landmark results in this direction, we refer to the works of
Lieb, Seiringer, Yngvason, and their collaborators 
which have initiated much of the current interest in the field, 
see \cite{ailisesoyn,lise,lisesoyn,liseyn} 
and the references therein. 

Another  main line of research focuses on the effective mean field dynamics of interacting Bose gases.
In the recent years, remarkable progress has been achieved in the 
mathematically rigorous derivation of
the nonlinear Schr\"{o}dinger (NLS) and nonlinear Hartree (NLH) equations
as the mean field limits of interacting Bose gases. 
For some recent fundamental results in this area, 
we refer to the works of Erd\"os, Schlein and Yau in \cite{esy1,esy2,ey},
and also \cite{klma, kiscst, rosc} and the references therein; see also
\cite{adgote,ansi,eesy,frgrsc,frknpi,frknsc,grmama,grma,he,sp}.

The strategy of \cite{esy1,esy2,ey} involves the following main steps, which are 
presented in more detail below: Based on the Schr\"odinger evolution
of the given $N$-body system
of bosons, one derives the associated BBGKY hierarchy of marginal 
density matrices. Subsequently, one takes the limit $N\rightarrow\infty$,
whereupon the BBGKY hierarchy tends to an infinite hierarchy of
marginal density matrices referred to as the Gross-Pitaevskii (GP) hierarchy.
Finally, one proves that for factorized initial conditions, the solutions
of the GP hierarchy are factorized and unique, and that the individual factors
satisfy the NLS or NLH, depending on the definition of the original $N$-body system.
In the work at hand, we will focus on the case linked to the NLS.

It is well known that for focusing $L^2$-critical and
$L^2$-supercritical NLS, negativity of the conserved energy implies blowup
of solutions in $H^1$. This is usually proven by use of energy conservation combined with
a virial identity, a method often referred to as Glassey's argument.
In this paper, we are especially interested in the phenomenon of blowup 
of solutions for the GP hierarchy without assuming factorization of the
initial conditions. More precisely, here we obtain an analogue of Glassey's argument
for the GP hierarchy, and thereby, we establish blowup of solutions to the GP hiearchy 
under the condition that the initial energy is negative. 

First, for the convenience of the reader, we outline below the 
main steps along which the defocusing cubic NLS is derived 
as the mean field limit for a gas of bosons with repelling pair interactions,
following \cite{esy1,esy2,ey}.  For repelling three body interactions leading to the
defocusing quintic NLS,
we refer to \cite{chpa}. We remark that it is currently not known how to obtain
analogous results for the case of attractive interactions.
\\

\noindent{\em (i) }\underline{\em From $N$-body Schr\"odinger to BBGKY.}
Let $\psi_{N}\in L^2(\R^{dN})$ denote the wave function describing $N$ bosons in $\R^d$. 
To account for the Bose-Einstein statistics, $\psi_{N}$ is invariant with respect to 
permutations  $\pi \in S_{N}$, which act by interchanging the particle variables,  
\begin{equation}\label{sym}
\psi_N(x_{\pi( 1)},x_{\pi (2)},...,x_{\pi (N)})=\psi_N(x_1, x_2,..., x_N) \,.
\end{equation} 
We denote $L_{s}^{2}(\Bbb R^{dN}):=\{\psi_N\in L^{2}(\Bbb R^{dN})| \, \psi_N {\rm \; satisfies \;}\eqref{sym}\}$. 
The dynamics of the system is determined by the Schr\"odinger equation
\begin{equation}\label{ham1}
i\partial_{t}\psi_{N}=H_{N}\psi_{N} \,.
\end{equation} 
The Hamiltonian $H_{N}$ is assumed to be a self-adjoint operator acting on the 
Hilbert space $L_{s}^2(\Bbb R^{dN})$, of the form  
\begin{equation}\label{ham2}
H_{N}=\sum_{j=1}^{N}(-\Delta_{x_{j}})+\frac1N\sum_{1\leq i<j \leq N}V_N(x_{i}-x_{j}),
\end{equation}
where $V_N(x)=N^{d\beta}V(N^\beta x)$ with 
$V\in W^{r,s}(\R^d)$ spherically symmetric, for some suitable $r$, $s$, 
and for $\beta\in(0,1)$ sufficiently small.

The limit $N\rightarrow\infty$ is obtained in the following manner.
One introduces the density matrix
$$\gamma_{N}(t, \ux_N, \ux_N')=\psi_{N}(t, \ux_N)\overline{\psi_{N}(t,\ux_N')}$$
where $\ux_N=(x_1,x_2,..., x_N)$ and $\ux_N'=(x_{1}^{\prime}, x_{2}^{\prime},..., x_{N}^{\prime})$.
Moreover, one introduces the associated sequence of 
$k$-particle marginal density matrices $\gamma_{N}^{(k)}(t)$, for $k=1,\dots,N$, 
as the partial trace of $\gamma_{N}$ 
over the degrees of freedom of the last $(N-k)$ particles,
$$\gamma_{N}^{(k)}=\Tr_{k+1, k+2,...,N}|\psi_{N}\rangle\langle\psi_{N}| \,.$$
Here, $\Tr_{k+1, k+2,...,N}$ denotes the partial trace with respect to the particles indexed
by $k+1, k+2,..., N$. 
Accordingly, $\gamma_{N}^{(k)}$ is defined as the non-negative trace class operator 
on $L_{s}^{2}(\Bbb R^{dk})$ with kernel given by
\eqn
	\gamma_{N}^{(k)}(\ux_k,\ux_k')
	&=& \int d\ux_{N-k} \gamma_{N}(\ux_k, \ux_{N-k};\ux_k', \ux_{N-k})
	\nonumber\\
	\label{reduced}
	&=&\int d\ux_{N-k}\overline{ \psi_{N}(\ux_k, \ux_{N-k}) }  \psi_{N}(\ux_k', \ux_{N-k}) \,.
\eeqn
It is clear from the definitions given above that $\gamma^{(k)}_N=\tr_{k+1}\gamma^{(k+1)}_N$, and that
$\Tr \gamma_{N}^{(k)}=\|\psi_N\|_{L_{s}^{2}(\Bbb R^{dN})}^2=1$ for all $N$, 
and all $k=1, 2, ..., N$. 

The time evolution of the density matrix $\gamma_{N}$ is determined by the Heisenberg equation
\begin{equation}\label{von}
i\partial_{t}\gamma_{N}(t)=[H_{N}, \gamma_N(t)] \, ,
\end{equation}
which is equivalent to
\eqn
	i\partial_{t}\gamma_{N}(t,\ux_N,\ux_N')
	&=&-(\Delta_{\ux_N}-\Delta_{ \ux_N' })\gamma_{N}(t,\ux_N, \ux_N') 
	\label{von2}\\
	&&+\frac{1}{N}\sum_{1\leq i<j \leq N}[V_N(x_i-x_j)-V_N(x_i^{\prime}-x_{j}^{\prime})]
	\gamma_{N}(t, \ux_N,\ux_N') \,, 
	\nonumber
\eeqn
expressed in terms of the associated integral kernel.
Accordingly, the $k$-particle marginals satisfy the BBGKY hierarchy 
\eqn\label{BBGKY}
	\lefteqn{
	i\partial_{t}\gamma^{(k)}(t,\ux_k;\ux_k')
	 \, = \, 
	-(\Delta_{\ux_k}-\Delta_{ \ux_k'})\gamma^{(k)}(t,\ux_k,\ux_k')
	}
	\nonumber\\
	&&
	+ \frac{1}{N}\sum_{1\leq i<j \leq k}[V_N(x_i-x_j)-V_N(x_i^{\prime}-x_{j}')]
	\gamma^{(k)}(t, \ux_{k};\ux_{k}') 
	\label{eq-bbgky-1}\\
	&&+\frac{N-k}{N}\sum_{i=1}^{k}\int dx_{k+1}[V_N(x_i-x_{k+1})-V_N(x_i^{\prime}-x_{k+1})]
	\label{eq-bbgky-2}\\
	&&\quad\quad\quad\quad\quad\quad\quad\quad\quad\quad\quad\quad\quad\quad\quad\quad
	\gamma^{(k+1)}(t, \ux_{k},x_{k+1};\ux_{k},x_{k+1}')
	\nonumber
\eeqn
where $\Delta_{\ux_k}:=\sum_{j=1}^{k}\Delta_{x_j}$, and similarly for $\Delta_{\ux_k'}$. 
We note that the number of terms in (\ref{eq-bbgky-1}) is $\approx \frac{k^2}{N}\rightarrow0$, 
and the number of terms in (\ref{eq-bbgky-2}) is $\frac{k(N-k)}{N}\rightarrow k$ 
as $N\rightarrow \infty$. Accordingly, for fixed $k$, (\ref{eq-bbgky-1}) disappears in the limit 
$N\rightarrow\infty$ described below, while (\ref{eq-bbgky-2}) survives.
\\

\noindent{\em (ii) }\underline{\em From BBGKY to GP.}
It is proven in \cite{esy1,esy2,ey} that, for a suitable topology on the space of marginal
density matrices, and  as $N\rightarrow\infty$, one can extract convergent subsequences
$\gamma^{(k)}_N\rightarrow\gamma^{(k)}$ for $k\in\N$, which satisfy
the infinite limiting hierarchy  
\eqn
	i\partial_{t}\gamma^{(k)}(t,\ux_k;\ux_k')
	&=&
 	- \, (\Delta_{\ux_k}-\Delta_{\ux_k'})\gamma^{(k)}(t,\ux_k;\ux_k')
	\label{GP}\\
	&&+ \, b_0 \sum_{j=1}^{k}   B_{j, k+1} 
	\gamma^{k+1}(t,\ux_k  ; \ux_k' )  \,, 
	\nonumber
\eeqn
which is referred to as the {\em Gross-Pitaevskii (GP) hierarchy}.
Here,
\eqn
	\lefteqn{
	(B_{j, k+1} \gamma^{k+1})(t,  \ux_k ; \ux_k' )
	}
	\nonumber\\
	&:=&
	\int dx_{k+1}dx_{k+1}'[\delta(x_j-x_{k+1})\delta(x_{j}-x_{k+1}^{\prime})-\delta(x_j^{\prime}-x_{k+1})
	\delta(x_{j}^{\prime}-x_{k+1}^{\prime})]
	\nonumber\\
	&&\quad\quad\quad\quad\quad\quad\quad\quad\quad\quad\quad\quad\quad\quad\quad\quad
	\gamma^{(k+1)}(t,\ux_k, x_{k+1};\ux_k', x'_{k+1}) \,,
	\nonumber 
\eeqn
and $b_0=\int V(x) dx$. 
The interaction term here is obtained from the limit of  (\ref{eq-bbgky-2}) as 
$N\rightarrow\infty$, using that $V_N(x)\rightarrow b_0\delta(x)$ weakly.
We will set $b_0=1$ in the sequel.
\\

\noindent{\em (iii) }\underline{\em NLS and factorized solutions of GP.}
The link between the original bosonic $N$-body system and solutions of the NLS is 
established as follows.
Given factorized  $k-$particle marginals
$$\gamma_0^{(k)}(\ux_k;\ux_k')=\prod_{j=1}^{k}\phi_0( x_{j})\overline{\phi_0( x_{j}^{\prime}})$$
at initial time $t=0$, with $\phi_0\in H^1(\R^d)$, one can easily verify that the 
solution of the GP hierarchy remains factorized for all $t\in I\subseteq\R$,
$$\gamma_0^{(k)}(t,\ux_k;\ux_k')=\prod_{j=1}^{k}\phi(t,x_{j})\overline{\phi(t,x_{j}^{\prime})} \,,$$
if $\phi(t)\in H^1(\R^d)$ solves the defocusing cubic NLS,
\eqn
	i\partial_t\phi \, = \, - \Delta_x \phi \, + \, |\phi|^2\phi\,,
\eeqn 
for $t\in I$, and $\phi(0)=\phi_0\in H^1(\R^d)$.  

Solutions of the GP hierarchy are studied in   spaces
of $k$-particle marginals with norms $\|\gamma^{(k)}\|_{H^1_k}^\sharp:= \Tr (S^{(k)}\gamma^{(k)})<\infty$
or $\|\gamma^{(k)}\|_{H^1_k}:= (\Tr (S^{(k)}\gamma^{(k)})^2)^{1/2}<\infty$
where $S^{(k)}:=\prod_{j=1}^k \langle\nabla_{x_j}\rangle\langle\nabla_{x_j'}\rangle$,
and $H^\alpha_k\equiv H^\alpha(\R^{dk}\times\R^{dk})$ for brevity.
While the existence of factorized solutions can be easily obtained, as
outlined above, the question remains whether solutions of the GP hierarchy
are also {\em unique}.

The proof of uniqueness of solutions of the GP hierarchy is the most difficult part in the 
program outlined above, and it was originally 
accomplished by Erd\"os, Schlein and Yau in \cite{esy1,esy2,ey} 
by use of sophisticated Feynman graph expansion methods.  
In \cite{klma} Klainerman and Machedon proposed an alternative method for proving uniqueness 
based on use of space-time bounds on the density matrices and introduction of an elegant 
``board game'' argument whose purpose is to organize the relevant combinatorics 
related to expressing solutions of the GP hierarchy using iterated Duhamel formulas. 
For the approach developed in  \cite{klma}, the authors assume that the a priori 
space-time bound 
\eqn \label{intro-KMbound} 
     \|B_{j;k+1} \gamma^{(k+1)}\|_{L^1_t\dot{H}^1_k} < C^k, 
\eeqn
holds, with $C$ independent of $k$. The authors of \cite{kiscst} 
proved that the latter is indeed satisfied for the cubic  case in $d=2$, 
based on energy conservation. 
\\

\noindent\underline{\em Non-factorized solutions of focusing and defocusing GP hierarchies.}
As mentioned above, it is currently only known how to obtain a GP hierarchy from the
$N\rightarrow\infty$ limit of a BBGKY hierarchy with  repulsive
interactions, but not for attractive interactions. 

However, in the work at hand, we will, similarly as in \cite{chpa2}, 
start directly from the level of the GP hierarchy, and 
allow ourselves to also discuss {\em attractive} interactions. 
Accordingly, we will refer to the corresponding GP hierarchies as {\em cubic},
{\em quintic}, {\em focusing}, or {\em defocusing GP hierarchies}, 
depending on the type of the NLS governing the solutions obtained from
factorized initial conditions.

Recently, in \cite{chpa2}, two of us analyzed the Cauchy problem 
for the cubic and quintic GP hierarchy in ${\mathbb R}^d$, $d\geq 1$ with focusing 
and defocusing interactions, and proved the existence and 
uniqueness of solutions to the GP hierarchy 
that satisfy the space-time bound \eqref{intro-KMbound} which was assumed in \cite{klma}.
As a key ingredient of the arguments in \cite{chpa2} a suitable
topology is introduced on the space of sequences of marginal density matrices,
\eqn\label{bigG}
	\Gspace \, = \, \{ \, \Gamma \, = \, ( \, \gamma^{(k)}(x_1,\dots,x_k;x_1',\dots,x_k') \, )_{k\in\N} 
	\, | \,
	\tr \gamma^{(k)} \, < \, \infty \, \} \,.
\eeqn 
It is determined by the generalized Sobolev norms
\eqn
	\| \, \Gamma \, \|_{\cH_\xi^\alpha} \, := \,
	\sum_{k\in\N} \xi^k \, \| \, \gamma^{(k)} \, \|_{H^\alpha_k} \,,
\eeqn
parametrized by $\xi>0$, and the spaces
$\cH_\xi^\alpha=\{ \, \Gamma\in\Gspace \,  | \, \| \, \Gamma \, \|_{\cH_\xi^\alpha} <\infty \, \}$  
were introduced.
The parameter $\xi>0$ is determined by the initial condition, and it sets the energy scale of 
a given Cauchy problem;
if $\Gamma\in\cH_\xi^\alpha$, then $\xi^{-1}$ is an upper bound on 
the typical $H^\alpha$-energy per particle.
The parameter $\alpha$ determines the regularity of the solution. 
In \cite{chpa2}, the local in time existence and uniqueness
of solutions is established
for cubic, quintic, focusing and defocusing GP hierarchies  in $\cH_\xi^\alpha$ 
for  $\alpha$ in a range depending on $d$, which satisfy a spacetime bound 
$\|\opB\Gamma\|_{L^1_{t\in I}\cH^{\alpha}_{\xi}}<C\|\Gamma_0\|_{\cH^\alpha_{\xi_0}}$ for some $0<\xi\leq\xi_0$ 
(here $\opB \Gamma:= ( \, B_{k+\frac{p}{2}} \gamma^{(k+\frac{p}{2})} \, )_{k\in\N}$). 
The precise statement and the associated consequences that we will use in this paper 
are presented in the next section. This result implies, in particular, 
\eqref{intro-KMbound}.
 
In this paper we study solutions of focusing GP hierarchies without any factorization 
condition, 
and especially establish the following results characterizing the blowup of solutions: 
\begin{itemize} 
\item[{\em (1)}]
For defocusing cubic GP hierarchies in $d=1,2,3$, and defocusing quintic GP 
hierarchies in $d=1,2$, 
which are obtained as limits of BBGKY hierarchies as outlined above,
it is possible to derive a priori bounds on $\|\gamma^{(k)}(t)\|_{L^\infty_t H^1_k}$
based on energy conservation in the $N$-particle Schr\"odinger system,
see \cite{esy1,esy2,ey}, and also \cite{chpa,kiscst}.
However, on the level of the GP hierarchy, no conserved energy functional has so far been known.
We identify 
an observable corresponding to the average energy per particle, 
and we prove that it is conserved. 
\item[{\em (2)}] Furthermore, we  prove the virial identity on the level of the GP hierarchy 
that enables us to obtain an analogue  of Glassey's argument 
from the analysis of focusing NLS equations. As a consequence, we prove that all solutions 
to the focusing GP hierarchy at the $L^2$-critical or $L^2$-supercritical level blow up in finite time
if the energy per particle in the initial condition is negative. 
\end{itemize}

\subsection*{Organization of the paper} 
In Section \ref{sec-def} we present the notation and the preliminaries. 
The main results of the paper are stated in 
Section \ref{sec-results}. In Section \ref{sec-energy} 
we identify the average energy per particle and prove that it is a conserved quantity.  
In Section \ref{sec-virial}, we derive a virial identity that enables us 
to prove an analogue of Glassey's blow-up argument familiar from the analysis of NLS. 
The analogue of Glassey's blowup argument is presented in Section \ref{sec-Glassey}.

\section{Definition of the model and preliminaries} 
\label{sec-def}

We introduce the space 
\eqn
	\Gspace \, := \, \bigoplus_{k=1}^\infty L^2(\R^{dk}\times\R^{dk})  
\eeqn
of sequences of density matrices
\eqn
	\Gamma \, := \, (\, \gamma^{(k)} \, )_{k\in\N}
\eeqn
where $\gamma^{(k)}\geq0$, $\tr\gamma^{(k)} =1$,
and where every $\gamma^{(k)}(\ux_k,\ux_k')$ is symmetric in all components of $\ux_k$,
and in all components of $\ux_k'$, respectively, i.e. 
\begin{equation}\label{symmetry}
	\gamma^{(k)}(x_{\pi (1)}, ...,x_{\pi (k)};x_{\pi'( 1)}^{\prime},
	 ...,x_{\pi'(k)}^{\prime})=\gamma^{(k)}( 	x_1, ...,x_{k};x_{1}^{\prime}, ...,x_{k}^{\prime})
\end{equation}
\\
holds for all $\pi,\pi'\in S_k$. 
\\
Moreover, the $k$-particle marginals are hermitean,
\\ 
\begin{equation}\label{conj}
\gamma^{(k)}(\ux_k;\ux_k')=\overline{\gamma^{(k)}(\ux_k';\ux_k) }.
\end{equation}

We call $\Gamma=(\gamma^{(k)})_{k\in\N}$  admissible if  
$\gamma^{(k)}=\tr_{k+1,\dots,k+\frac p2}\gamma^{(k+\frac p2)}$, that is,
\eqn
	\lefteqn{
	\gamma^{(k)}(\ux_k;\ux_k') 
	}
	\\
	&&\, = \, \int dx_{k+1} \cdots dx_{k+\frac p2}
	\, \gamma^{(k+\frac{p}{2})}(\ux_{k},x_{k+1},\dots,x_{k+\frac p2};\ux_k',x_{k+1},\dots,x_{k+\frac p2}) 
	\nonumber
\eeqn  
for all $k\in\N$.

We will use the following convention for the Fourier transform,
$$
	\gamma(\ux_k;\ux_k^{\prime})=\int d\uu_kd\uu_k^{\prime}
	e^{i\uu_k\ux_k-i\uu_k^{\prime}\ux_k^{\prime}}\widehat{\gamma}(\uu_k;\uu_k^{\prime}).
$$

Let $0<\xi<1$.
We define
\eqn
	\cH_\xi^\alpha \, := \, \Big\{ \, \Gamma \, \in \, \Gspace \, \Big| \, \|\Gamma\|_{\cH_\xi^\alpha} < \, \infty \, \Big\}
\eeqn
where
\eqn
	\|\Gamma\|_{\cH_\xi^\alpha} \, = \, \sum_{k=1}^\infty \xi^{ k} 
	\| \,  \gamma^{(k)} \, \|_{H^\alpha(\R^{dk}\times\R^{dk})} \,,
\eeqn
with
\eqn\label{def-Sobnorms-1}
	\| \, \gamma^{(k)} \, \|_{H^\alpha(\R^{dk}\times\R^{dk})} \, = \,
	\| \, S^{(k,\alpha)} \, \gamma^{(k)} \, \|_{L^2(\R^{dk}\times\R^{dk})}  \,,
\eeqn
and $S^{(k,\alpha)}:=\prod_{j=1}^k\langle\nabla_{x_j}\rangle^\alpha\langle\nabla_{x_j'}\rangle^\alpha$. Clearly, $\cH_\xi^\alpha$ is a Banach space.
Similar spaces are used in  the isospectral renormalization group analysis of spectral
problems in quantum field theory, \cite{bcfs}.

Next, we define the cubic, quintic, focusing, and defocusing GP hierarchies.
Let $p\in\{2,4\}$.  
The $p$-GP (Gross-Pitaevskii) hierarchy is given by
\eqn \label{eq-def-b0-2}
	i\partial_t \gamma^{(k)} \, = \, \sum_{j=1}^k [-\Delta_{x_j},\gamma^{(k)}]   
	\, + \,  \mu B_{k+\frac p2} \gamma^{(k+\frac p2)}
\eeqn
in $d$ dimensions, for $k\in\N$. Here,
\eqn
	\lefteqn{
	\left(B_{k+\frac p2}\gamma^{(k+\frac{p}{2})}\right)(t,x_1,\dots,x_k;x_1',\dots,x_k')
	}
	\\
	&& := \,
	\sum_{j=1}^k \left(B_{j;k+1,\dots,k+\frac p2}\gamma^{(k+\frac{p}{2})}\right)(t,x_1,\dots,x_k;x_1',\dots,x_k')
	\nonumber\\
	&& := \,
        \sum_{j=1}^k \big[\, \left(B^1_{j;k+1,\dots,k+\frac p2}\gamma^{(k+\frac{p}{2})}\right)(t,x_1,\dots,x_k;x_1',\dots,x_k')
	\nonumber\\ 
        &&\quad\quad\quad\quad 
        - \left(B^2_{j;k+1,\dots,k+\frac p2}\gamma^{(k+\frac{p}{2})}\right)(t,x_1,\dots,x_k;x_1',\dots,x_k')\, \big],
        \nonumber
\eeqn
where  
\begin{align*} 
& \left(B^1_{j;k+1,\dots,k+\frac p2}\gamma^{(k+\frac{p}{2})}\right)
(t,x_1,\dots,x_k;x_1',\dots,x_k') \\
& \quad \quad = \int dx_{k+1}\cdots dx_{k+\frac p2} dx_{k+1}'\cdots dx_{k+\frac p2}' \\
& \quad\quad\quad\quad 
	\prod_{\ell=k+1}^{k+\frac p2} \delta(x_j-x_{\ell})\delta(x_j-x_{\ell}' )
        \gamma^{(k+\frac p2)}(t,x_1,\dots,x_{k+\frac p2};x_1',\dots,x_{k+\frac p2}'),
\end{align*} 
and 
\begin{align*} 
& \left(B^2_{j;k+1,\dots,k+\frac p2}\gamma^{(k+\frac{p}{2})}\right)
(t,x_1,\dots,x_k;x_1',\dots,x_k') \\
& \quad \quad = \int dx_{k+1}\cdots dx_{k+\frac p2} dx_{k+1}'\cdots dx_{k+\frac p2}' \\
& \quad\quad\quad\quad 
	\prod_{\ell=k+1}^{k+\frac p2} \delta(x'_j-x_{\ell})\delta(x'_j-x_{\ell}' )
        \gamma^{(k+\frac p2)}(t,x_1,\dots,x_{k+\frac p2};x_1',\dots,x_{k+\frac p2}').
\end{align*} 
The operator $B_{k+\frac p2}\gamma^{(k+\frac{p}{2})}$
accounts for  $\frac p2+1$-body interactions between the Bose particles.
We note that for factorized solutions, the corresponding 1-particle wave function satisfies the
$p$-NLS $i\partial_t\phi=-\Delta\phi+\mu|\phi|^p\phi$.

We refer to (\ref{eq-def-b0-2}) as the {\em cubic GP hierarchy} if $p=2$,
and as the {\em quintic GP hierarchy} if $p=4$. 
Also we denote the $L^2$-critical exponent by $p_{L^2} = \frac 4d$
and refer to  (\ref{eq-def-b0-2})  as a {\em $L^2$-critical GP hierarchy} if $p=p_{L^2}$ 
and as a {\em $L^2$-supercritical GP hierarchy} if $p > p_{L^2}$. 
Moreover, for $\mu=1$ or $\mu=-1$ we refer to the GP hierarchies as being  
defocusing or focusing, respectively. 

To obtain the blow-up property, we need a result providing a blow-up alternative. 
This is a usually obtained as a byproduct of the local theory. 
In the context of the local theory developed in \cite{chpa2}, 
we recall the following two theorems: Theorem \ref{thm-main-0}
which establishes the local well-posedness of the GP equation 
and Theorem \ref{thm-blowuprate-L2crit-1} that gives lower bounds on the blow-up rate. 
In order to state these two theorems, we recall that in \cite{chpa2} 
the GP hieararchy was rewritten in the following way: 
\eqn \label{chpa2-pGP}
        i\partial_t \Gamma \, + \, \opDelta_\pm \Gamma \, = \, \mu \opB \Gamma \,,
\eeqn 
where
$$
	\opDelta_\pm \Gamma \, := \, ( \, \Delta^{(k)}_\pm \gamma^{(k)} \, )_{k\in\N}
        \; \; \; \; \mbox{ with } \; \; \; \; \Delta_{\pm}^{(k)} \, = \, \Delta_{\ux_k} - \Delta_{\ux'_k},
$$
and 
\eqn \label{chpa2-B} 
	\opB \Gamma \, := \, ( \, B_{k+\frac{p}{2}} \gamma^{(k+\frac{p}{2})} \, )_{k\in\N} \,.
\eeqn

Also the following set $\alphaset(d,p)$ was introduced in \cite{chpa2}, for $p=2,4$ 
and $d\geq1$, 
\eqn\label{eq-alphaset-def-1}
	\alphaset(d,p) \, = \, \left\{
	\begin{array}{cc}
	(\frac12,\infty) & {\rm if} \; d=1 \\ 
	(\frac d2-\frac{1}{2(p-1)}, \infty) & {\rm if} \; d\geq2 \; {\rm and} \; (d,p)\neq(3,2)\\
	\big[1,\infty) & {\rm if} \; (d,p)=(3,2)
	\end{array}
	\right.
\eeqn 
which we will use to account for the regularity of solutions.

Now we recall the local well-posedness theorem proved in \cite{chpa2}. 
\begin{theorem}\label{thm-main-0}
Let $\xi_1>0$.  
Assume that $\alpha\in\alphaset(d,p)$ where $d\geq1$ and $p\in\{2,4\}$, and $0<\eta<1$ sufficiently small.
Then, the following hold.
\begin{itemize}
\item[(i)]
For every $\Gamma_0\in\cH_{{\xi_1}}^\alpha$, there
exist constants $T>0$ and $0<\xi_2\leq\xi_1$ 
such that the following holds. There exists a unique
solution $\Gamma(t)$ in the space
$$
	\{\Gamma\in L^\infty_{t\in I}\cH_{\xi_2}^\alpha
	\, | \, \|\opB\Gamma\|_{L^1_{t\in I}\cH_{\xi_2}^\alpha}<\infty\} \,.
$$  
where $I:=[0,T]$.
In particular this solution satisfies the Strichartz-type bound
\eqn\label{eqn-BGamma-spacetime-bd-0}
	\|\opB\Gamma\|_{L^1_{t\in I}\cH_{\xi_2}^\alpha}
	&\leq&C(T,d,p,\xi_1,\xi_2) \,
	\|\Gamma_0\|_{\cH_{\xi_1}^\alpha}  \,.
\eeqn
$\;$
\\
\item[(ii)]
The uniqueness of solutions in the space 
$L^\infty_{t\in I}\cH_{\xi_2}^\alpha$ is characterized as follows.
Given $\Gamma_0\in\cH_{{\xi_1}}^\alpha$, assume that there are constants $T>0$ 
and $0<\xi_2\leq\xi_1$ such that there exists a  
solution $\Gamma(t)$ of the $p$-GP hierarchy (\ref{eq-def-b0-2})
in the space $L^\infty_{t\in I}\cH_{\xi_2}^\alpha$. 
\\
\\
Then, this solution 
is {\em unique} in $L^\infty_{t\in I}\cH_{\xi_2}^\alpha$ if and only if 
$\| \, \opB \Gamma \, \|_{L^1_{t\in I}\cH_{\xi}^\alpha}<\infty$ 
holds for some $\xi>0$. 
\end{itemize}
\end{theorem}

\begin{definition}\label{def-blowup-1}
We say that a solution $\Gamma(t)$ of the GP hierarchy blows up in finite time 
with respect to $H^\alpha$ if 
there exists $T^*<\infty$  such that the following holds:
For every $\xi>0$ there exists $T_{\xi,\Gamma}^*<T^*$ such that
$\|\Gamma(t)\|_{\cH_{\xi}^\alpha}\rightarrow\infty$ as $t\nearrow T^{*}_{\xi,\Gamma}$.
Moreover, $T_{\xi,\Gamma}^*\nearrow T^*$ as $\xi\rightarrow0$. 
\end{definition}

For the study of blowup solutions, it is convenient to introduce the following quantity.

\begin{definition}\label{def-AvHLp-1}
We refer to
\eqn
	\Av_{H^\alpha}(\Gamma ) \, := \, \Big[ \, \sup\big\{ \, \xi >0 \, \big| \,
	\| \, \Gamma \, \|_{\cH_\xi^\alpha} <\infty\, \big\} \, \Big]^{-1} \,,
\eeqn
\eqn
	\Av_{L^r}(\Gamma ) \, := \, \Big[ \, \sup\big\{ \, \xi >0 \, \big| \,
	\| \, \Gamma \, \|_{\cL_\xi^r} <\infty\, \big\} \, \Big]^{-1} \,,
\eeqn
respectively, as the typical (or average) $H^\alpha$-energy and the typical $L^r$-norm per particle. 
\end{definition}

We then have the following characterization of blowup.

\begin{lemma}
Blowup in finite time of $\Gamma(t)$ with respect to $H^\alpha$ as $t\nearrow T^*$,
as characterized in Definition \ref{def-blowup-1},
is equivalent to the statement that $\Av_{H^\alpha}(\Gamma(t))\rightarrow\infty$
as $t\nearrow T^*$ (and similarly for $L^r$). 
\end{lemma}

\prf
Clearly, $\Av_{H^\alpha}(\Gamma )$ is the reciprocal of the convergence radius of
$\|\Gamma\|_{\cH^\alpha_\xi}$ as a power series in $\xi$. Accordingly, 
$\|\Gamma\|_{\cH^\alpha_\xi}<\infty$ for $\xi<\Av_{H^\alpha}(\Gamma )^{-1}$,
and $\|\Gamma\|_{\cH^\alpha_\xi}=\infty$ for $\xi\geq\Av_{H^\alpha}(\Gamma )^{-1}$.

Blowup in finite time of $\Gamma(t)$ in $H^\alpha$ as $t\nearrow T^*$,
as characterized in Definition \ref{def-blowup-1},
is equivalent to the statement that the convergence radius of 
$\|\Gamma(t)\|_{\cH^\alpha_\xi}$, as a power series in $\xi$, tends to zero as $t\nearrow T^*$.
Thus, in turn, blowup in finite time of $\Gamma(t)$ with respect to $H^\alpha$
is equivalent to the statement that $\Av_{H^\alpha}(\Gamma(t))\rightarrow\infty$
as $t\nearrow T^*$. 
\endprf

The following theorem from \cite{chpa2} gives lower bounds on the blow-up rate. 
\begin{theorem}
\label{thm-blowuprate-L2crit-1}
Assume that $\Gamma(t)$ is a solution of the (cubic $p=2$ or $p=4$ quintic) $p$-GP hierarchy
with initial condition $\Gamma(t_0)=\Gamma_{0}\in\cH_{\xi}^\alpha$,
for some $\xi>0$, which blows up in finite time. Then, the following
lower bounds on the blowup rate hold:
\begin{enumerate}
\item[($a$)] 
Assume that $\frac4d\leq p < \frac{4}{d-2\alpha}$. Then,
\eqn
	( \, \Av_{H^\alpha}(\Gamma(t)) \, )^{\frac12}  
	\, > \, \frac{C}{|T^*-t|^{(2\alpha-d+\frac4p)/4}} \,.
\eeqn 
Thus specifically, for the cubic GP hierarchy in $d=2$, and for the quintic GP hierarchy in $d=1$,
\eqn
	( \, \Av_{H^1}(\Gamma(t )) \, )^{\frac12} \, \geq \, \frac{C}{|t -T^*|^{\frac12}} \,,
\eeqn 
with respect to the Sobolev spaces $H^\alpha$, $\cH_\xi^\alpha$.

\item[($b$)] 
\eqn
	( \, \Av_{L^r}(\Gamma(t )) \, )^{\frac12} \, \geq \, 
        \frac{C}{|t -T^*|^{\frac{1}{p}-\frac{d}{2r}}} \,, \mbox{ for } \frac{pd}{2} < r.
\eeqn 
\end{enumerate} 
\end{theorem}

\begin{remark}
We note that in the factorized case, the above 
lower bounds on the blow-up rate coincide with the known
lower bounds on the blow-up rate for solutions to the NLS (see, for example, \cite{ca}).
\end{remark} 

We note that 
\eqn
	\Gamma \, = \, ( \, | \, \phi \, \rangle \langle \, \phi \, |^{\otimes k} \, )_{k\in\N}
	\; \; \; \Rightarrow \; \; \; 
 	\Av_{H^\alpha}(\Gamma ) \, = \, \|\phi\|_{H^\alpha}^2 \mbox{ and } 
	\Av_{L^r}(\Gamma ) \, = \, \|\phi\|_{L^r}^2
\eeqn
in the factorized case.

The fact that $\Gamma\in\cH_\xi^\alpha$ means that
the typical energy per particle is bounded by $\Av_{H^\alpha}(\Gamma )<\xi^{-1}$.
Therefore, the parameter $\xi$ determines the $H^\alpha$-energy scale in the problem.
While solutions with a bounded $H^\alpha$-energy remain in the same $\cH_\xi^\alpha$
for some sufficiently small $\xi>0$,
blowup solutions undergo transitions
$\cH_{\xi_1}^\alpha\rightarrow\cH_{\xi_2}^\alpha\rightarrow\cH_{\xi_3}^\alpha\rightarrow\cdots$
where the sequence $\xi_1>\xi_2>\cdots$ converges to zero as $t\rightarrow T^*$.

We emphasize again that $(\Av_{N}(\Gamma ))^{-1}$ is the convergence radius of $\|\Gamma\|_{\cN_\xi}$
as a power series in $\xi$, for the norms $N=H^\alpha,L^r$ and 
$\cN_\xi=\cH^\alpha_\xi ,\cL^r_\xi$, respectively.

\section{Statement of the main results} 
\label{sec-results}

The following two Theorems are the main results of this paper. 
First we prove energy conservation per particle for solutions 
$\Gamma(t)$ of the $p$-GP hierarchy. More precisely, 
\begin{theorem}\label{thm-main-enconserv-1}
Let $0<\xi<1$. Assume that $\Gamma(t)\in\cH_\xi^\alpha$, with $\alpha\geq1$, is a 
solution of the focusing ($\mu=-1$) or defocusing ($\mu=1$) 
$p$-GP hierarchy with initial condition $\Gamma_0\in\cH_\xi^\alpha$.  
Then, the following hold. 
Let
\eqn\label{eq-Ek-def-1}
	E_k( \, \Gamma(t) \, ) & := & \frac12\tr( \, \sum_{j=1}^k(-\Delta_{x_j}) \gamma^{(k)}(t)  \, ) 
	\\
	&&\quad\quad\quad\quad
	\, + \, \frac{\mu}{p+2} \tr( \,  \sum_{j=1}^k B_{j;k+1,\dots,k+\frac p2}^1 \, \gamma^{(k+\frac p2)}(t) \, ).  
	\nonumber
\eeqn
Then, the quantity
\eqn
	\En_{\xi}( \, \Gamma(t) \, )
	\, := \, \sum_{k\geq1}\xi^k \, E_k( \, \Gamma(t) \, )
\eeqn  
is conserved, and in particular,
\eqn\label{eq-Enxi-def-sum-1}
	\En_{\xi}( \, \Gamma(t) \, )
	\, = \, \big( \sum_{k\geq1}k \xi^k \, \big) \, E_1( \, \Gamma(t) \, ) \,,
\eeqn 
where $\sum_{k\geq1}k \xi^k<\infty$ for any $0<\xi<1$. 
\end{theorem}

We recall that $\tr(A)$ means integration of the kernel $A(x,x')$ against the measure
$\int dx dx'\delta(x-x')$.
 We note that for factorized states $\Gamma(t)=(|\phi(t)\rangle\langle\phi(t)|^{\otimes k})_{k\in\N}$,
one finds
\eqn
	 E_1( \, \Gamma(t) \, )  \, = \, 
  \frac12\|\nabla\phi(t)\|_{L^2}^2+\frac{\mu}{p+2}\|\phi(t)\|_{L^{p+2}}^{p+2} \,,
\eeqn
which is the usual expression for the conserved energy for solutions of the NLS 
$i\partial_t\phi+\Delta \phi+\mu|\phi|^p\phi=0$.

\begin{theorem} \label{blowup-sufcond}
Let $p \geq p_{L^2}$. 
Assume that $\Gamma(t) = (\, \gamma^{(k)}(t) \,)_{k\in\N}$
solves the focusing (i.e., $\mu=-1$) 
$p$-GP hierarchy with initial condition $\Gamma(0)\in\cH_\xi^1$ for
some $0<\xi<1$, with $\tr( \, x^2 \gamma^{(1)}(0) \, ) < \infty$.
If  $ E_1( \, \Gamma(0) \, ) < 0$, then there exists $T^*<\infty$ 
such that $\Av_{H^1}(\Gamma(t))\rightarrow\infty$ as $t\nearrow T^*$. 
\end{theorem}

\begin{remark}
We note that Theorem \ref{blowup-sufcond} is proved under the assumption that 
$\tr( \, x^2 \gamma^{(1)}(0) \, ) < \infty$, which is analogous to the finite 
variance assumption in the case of Glassey's blow-up argument for the NLS (see, 
e.g. \cite{ca} Theorem 6.5.4).    
\end{remark}

As a motivation for the proofs presented below, we briefly recall the application of
Glassey's argument in the case of an $L^2$-critical or supercritical focusing NLS.
We consider a solution of $i\partial_t\phi=-\Delta\phi-|\phi|^p\phi$ with
$\phi(0)=\phi_0\in H^1(\R^d)$ and $p\geq p_{L^2}=\frac4d$,
such that the conserved energy satisfies
$E[\phi(t)]:=\frac12\|\nabla\phi(t)\|_{L^2}^2-\frac{1}{p+2}\|\phi(t)\|_{L^{p+2}}^{p+2}
=E[\phi_0]<0$.
Moreover, we assume that $\| \, |x| \phi_0 \, \|_{L^2}<\infty$. 
Then, one considers the quantity $V(t):=\langle\phi(t),x^2\phi(t)\rangle$, which is shown to
satisfy the virial identity  
\eqn\label{eq-NLS-virial-1}
  \partial_t^2 V(t) \, = \, 
  16 E[\phi_0]  \, - \, 4d \,  \frac{p - p_{L^2}}{p+2} \|\phi(t)\|_{L^{p+2}}^{p+2} \,.
\eeqn
Hence, if $E[\phi_0]<0$, and $p\geq p_{L^2}$, there exists a finite time $T^*$ such that the positive quantity 
$V(t)\searrow0$ as $t\nearrow T^*$. Accordingly, this  implies that
$\|\phi(t)\|_{H^1(\R^d)}\nearrow\infty$ as $t\nearrow T^*$ (for more
details, see Section \ref{sec-Glassey}). This phenomenon is referred to as
negative energy blowup in finite time for the NLS.
In the sequel, we will prove analogues of these arguments for the GP hierarchy.

\section{Conservation of energy}
\label{sec-energy}

In this section, we prove Theorem \ref{thm-main-enconserv-1}.
We first demonstrate the proof for the cubic case, $p=2$.
To begin with, we note that
\eqn
	E_k( \, \Gamma(t) \, ) \, = \, k E_1( \, \Gamma(t) \, )
\eeqn
for $\Gamma(t)=(\gamma^{(k)}(t))_{k\in\N}$ a solution of the cubic GP hierarchy,
where by definition, all $\gamma^{(k)}$'s are admissible.
To prove this, we note that  (\ref{eq-Ek-def-1}) can be written as
\eqn
	E_k( \, \Gamma(t) \, ) &  = & \sum_{j=1}^k\Big[ \, \frac12\tr( \,(-\Delta_{x_j}) \gamma^{(k)}(t)  \, ) 
	\nonumber\\
	&&\quad\quad\quad\quad
	\, + \, \frac{\mu}{4} \tr( \,   B_{j,k+1}^1 \, \gamma^{(k+1)}(t) \, )  \, \Big] \,,
\eeqn
where each of the terms in the sum equals the one obtained for $j=1$, 
by symmetry of $\gamma^{(k)}$ and $\gamma^{(k+1)}$
with respect to their variables. 
We present the detailed calculation for the interaction term, and note that the calculation 
for the kinetic energy term is similar. Consider $1 \leq i<j \leq k$. We have that
\eqn
	\lefteqn{
	(B_{j,k+1}^{1}\gamma^{(k+1)})(x_{1}, x_2,...,x_k;x_{1}^{\prime}, 
	x_2^{\prime},...,x_k^{\prime})
	}
	\\
 	&=& \gamma^{(k+1)}(x_{1}, x_2,...,x_i,...,x_j,...,x_k, x_{j};x_{1}^{\prime}, 		
	x_2^{\prime},...,x_{i}^{\prime},...,x_{j}^{\prime},...,x_k^{\prime}, x_{j}) \,,
	\nonumber
\eeqn
and
\eqn
	\lefteqn{
	(B_{i,k+1}^{1}\gamma^{(k+1)})(x_{1}, x_2,...,x_k;x_{1}^{\prime}, 
	x_2^{\prime},...,x_k^{\prime})
	}
	\\
	&=&\gamma^{(k+1)}(x_{1},  ...,x_i,...,x_j,...,x_k, x_{i};x_{1}^{\prime}, 	
	...,x_{i}^{\prime},...,x_{j}^{\prime},...,x_k^{\prime}, x_{i}) \, .
	\nonumber
\eeqn
Thus 
\eqn
	\lefteqn{
	\tr(B_{j,k+1}^{1}\gamma^{(k+1)})
	}
	\\
	&=&
	\int\gamma^{(k+1)}(x_{1},  ...,x_i,...,x_j,...,x_k, x_{j};x_{1},  ...x_{i},...,
	x_{j},...,x_k, x_{j})dx_1dx_2...dx_k \,.
	\nonumber
\eeqn
By the symmetry of $\gamma^{(k+1)}(\ux_{k+1};\ux_{k+1}')$ with respect to the components of
$\ux_{k+1}$ and $\ux_{k+1}'$, respectively,
$$\tr(B_{j,k+1}^{1}\gamma^{(k+1)})=\tr(B_{i,k+1}^{1}\gamma^{(k+1)})$$
for all $i,j\in\{1,\dots,k\}$.
Thus,
\eqn
	E_k( \, \Gamma(t) \, ) &  = &  k\Big[ \, \frac12\tr( \,(-\Delta_{x_1}) \gamma^{(k)}(t)  \, ) 
	\nonumber\\
	&&\quad\quad\quad\quad
	\, + \, \frac{\mu}{4} \tr( \,   B_{1,k+1}^1 \, \gamma^{(k+1)}(t) \, )  \, \Big] \,
\eeqn
follows.

We can go one step further and note that
\eqn
	\lefteqn{
	(B_{1,k+1}^{1}\gamma^{(k+1)})(x_{1}, x_2,...,x_k;x_{1}^{\prime}, 
	x_2^{\prime},..., x_k^{\prime})  
	}
	\nonumber\\
	&=&\gamma^{(k+1)}(x_{1}, x_2,...,x_k, x_{1};x_{1}^{\prime}, x_2^{\prime},..., x_k^{\prime}, x_{1}).
\eeqn
Then
\eqn
	\tr(B_{1,k+1}^{1}\gamma^{(k+1)})
	& = &
	\int\gamma^{k+1}(x_{1}, x_2,...,x_k, x_{1};x_{1}, x_2,...,x_k, x_{1})dx_1dx_2...dx_k
	\quad\quad \nonumber
	\\
	&=&\int\gamma^{(k+1)}(x_{1}, x_1,x_2,...,x_k;x_{1}, x_1, x_2,...,x_k)dx_1dx_2...dx_k
	\nonumber
\eeqn
where in the last equality we used symmetry of $\gamma^{(k+1)}$. On the other hand,
\eqn
	(B_{1,2}^{1}\gamma^{(2)})(x_{1};x_{1}^{\prime})
	&=&\gamma^{(2)}(x_1, x_1; x_1^{\prime}, x_1) 
	\\
	&=&\int\gamma^{(k+1)}(x_{1}, x_1, x_2,...,x_k ; x^{\prime}_{1}, x_1, x_2,...,x_k)dx_2...dx_k
	\nonumber
\eeqn
by repeated use of the admissibility of $\gamma^{(j)}$, for $j=2,\dots,k+1$.
Thus,
$$\tr(B_{1,2}^{1}\gamma^{(2)})=\int\gamma^{(k+1)}(x_{1}, x_1,x_2,...,x_k;x_{1}, x_1, x_2,...,x_k)dx_1dx_2...dx_k$$
and  
$$\tr(B_{1,k+1}^{1}\gamma^{(k+1)})=\tr(B_{1,2}^{1}\gamma^{(2)})$$
follows. Therefore,
\eqn
	E_k( \, \Gamma(t) \, ) &  = &  k\Big[ \, \frac12\tr( \,(-\Delta_{x_1}) \gamma^{(1)}(t)  \, ) 
	\, + \, \frac{\mu}{4} \tr( \,   B_{1,2}^1 \, \gamma^{(2)}(t) \, )  \, \Big] 
	\nonumber\\
	& = & k E_1( \, \Gamma(t) \, ) \,,
\eeqn
as claimed.

The fact that \eqref{eq-Enxi-def-sum-1} then follows is evident.

Next, we verify that $E_1( \, \Gamma(t) \, )$ is a conserved quantity, which means that 
$E_1( \, \Gamma(t) \, )=E_1( \, \Gamma(0) \, )$ for all $t\in\R$.

For the proof, the following auxiliary identities are very useful:
\eqn\label{eq-trderA-id-aux-1}
	\lefteqn{
	\int dxdx'\delta(x-x')
	\nabla_{x}\cdot\nabla_{x'} A(x;x') 
	}
	\nonumber\\
	& = &
	\int dxdx'\delta(x-x')
	\Delta_{x}A(x;x') 
	\label{aux-deltax}\\
	&=&\int dxdx'\delta(x-x')
	\Delta_{x'}A(x;x') \,. \label{aux-deltaxprim}
\eeqn
To prove \eqref{aux-deltax}, we note that
\eqn
	\lefteqn{
	\int dxdx'\delta(x-x')
	\nabla_{x}\cdot\nabla_{x'} A(x;x') 
	}
	\nonumber\\
	& = &
	-\int du du' \int dxdx'\delta(x-x') \, e^{iux-iu'x'}\,
	u\cdot u' \widehat A(u;u')
	\nonumber\\
	& = &
	-\int du du' \delta(u-u') \,
	u\cdot u' \widehat A(u;u') 
	\nonumber\\
	&=&
	-\int du du' \delta(u-u') \,
	u^2 \widehat A(u;u')
	\nonumber\\
	& = &
	-\int du du' \int dxdx'\delta(x-x') \, e^{iux-iu'x'}\,
	u^2 \widehat A(u;u')
	\nonumber\\
	&=&\int dx dx' \delta(x-x') \,
	\Delta_x \widehat A(x;x') \,.
\eeqn
The equality \eqref{aux-deltaxprim} can be proved in a similar way.

We now return to the proof of $E_1( \, \Gamma(t) \, )=E_1( \, \Gamma(0) \, )$.
For $k=1$, we consider $\gamma^{(1)}(x,x')$ where
\eqn
	E_1( \, \Gamma(t) \, ) & = &  -\frac12\tr( \, \nabla_{x}\cdot\nabla_{x'}\gamma^{(1)} \, )  
	\\
	&&
	\, + \, \frac{\mu}{4}\int dx_1dx_2dx_1'dx_2' \, 
	\delta(x_1=x_2=x_1'=x_2') \, \gamma^{(2)}(x_1,x_2;x_1',x_2') 
	\nonumber
\eeqn
in symmetrized form. Here, we have introduced the shorthand notation
\eqn
	 \delta(x_1=x_2=x_1'=x_2')  \, := \, \delta(x_1-x_2)\delta(x_1-x_2')\delta(x_2-x_2') \,.
\eeqn
Clearly,
\eqn
	i\partial_tE_1( \, \Gamma(t) \, ) \, = \, (I) \, + \, (II) \, + \, (III) \, + \, (IV)
\eeqn
where
\eqn
	(I) \, := \, \frac12\tr( \,  \nabla_{x}\cdot\nabla_{x'}(\Delta_{x}-\Delta_{x'})\gamma^{(1)} \, )    \,.
\eeqn
\eqn
	(II) \, := \, -  \frac\mu2\tr( \, \nabla_{x}\cdot\nabla_{x'}(B_{1,2}\gamma^{(2)}) \, )  
\eeqn
\eqn
	(III) & := &- \frac{\mu}{4}\int dx_1dx_2dx_1'dx_2' \, \delta(x_1=x_2=x_1'=x_2') \, 
	\\
	&&\quad\quad\quad\quad\quad\quad
	(\Delta_{x_1}+\Delta_{x_2}-\Delta_{x_1'}-\Delta_{x_2'})\gamma^{(2)}(x_1,x_2;x_1',x_2')
	\nonumber
\eeqn
\eqn
	(IV) & :=  &\frac{\mu}{4}\int dx_1dx_2dx_1'dx_2' \, \delta(x_1=x_2=x_1'=x_2') \, 
	\nonumber\\
	&&\quad\quad\quad\quad\quad\quad
	(B_3\gamma^{(3)})(x_1,x_2;x_1',x_2') \,.
\eeqn

\noindent\underline{\em The term $(I)$.}
We claim that $(I)=0$.
We note that
\eqn
	\lefteqn{
	\tr( \, \nabla_{x}\cdot\nabla_{x'}(\Delta_{x}-\Delta_{x'})\gamma^{(1)} \, ) 
	}
	\nonumber\\
	&=&\int du du' \, \int dx dx'\delta(x-x') \, u\cdot u'(u^2-(u')^2) \, e^{iux-iu'x'} \, \widehat\gamma^{(1)}(u;u')
	\nonumber\\
	&=&\int du du' \, \delta(u-u') \, u\cdot u'(u^2-(u')^2) \, \widehat\gamma^{(1)}(u;u')
	\nonumber\\
	&=&0 \,.
\eeqn
This proves the claim.
\\

\noindent\underline{\em The term $(IV)$.}
We claim that $(IV)$ also vanishes. Indeed,
\eqn 
	(IV) & =  &\frac{\mu}{4}\int dx_1dx_2dx_1'dx_2' \, \delta(x_1=x_2=x_1'=x_2') \, 
	\nonumber\\
	&&\quad\quad\quad\quad\quad\quad
	(B_3\gamma^{(3)})(x_1,x_2;x_1',x_2')  
	\nonumber\\
	 & =  &\sum_{j=1,2}\frac{\mu}{4}\int dx_1dx_2dx_1'dx_2' \, \delta(x_1=x_2=x_1'=x_2') \, 
	\nonumber\\
	&&\quad\quad\quad\quad\quad\quad
	\Big[ \, \gamma^{(3)}(x_1,x_2,x_1;x_1',x_2',x_1)- \gamma^{(3)}(x_1,x_2,x_1';x_1',x_2',x_1')
	\nonumber\\
	&&\quad\quad\quad\quad\quad\quad
	+ \, \gamma^{(3)}(x_1,x_2,x_2;x_1',x_2',x_2)- \gamma^{(3)}(x_1,x_2,x_2';x_1',x_2',x_2') \, \Big] 
	\nonumber\\ 
	 & =  &2 \, \frac{\mu}{4}\int dx 
	\Big[ \, \gamma^{(3)}(x,x,x;x,x,x)- \gamma^{(3)}(x,x,x;x,x,x) \, \Big]
	\nonumber\\
	& = & 0 \,.
\eeqn
This proves the claim.
\\

\noindent\underline{\em The term $(III)$.}
By symmetry of $\gamma^{(2)}(x_1,x_2;x_{1}',x_{2}')$ in $(x_1,x_2)$, and in $(x_{1}',x_{2}')$, we find
\eqn
	(III) & = &-2 \, \frac{\mu}{4}\int dx_1dx_2dx_1'dx_2' \, \delta(x_1=x_2=x_1'=x_2') \, 
	\nonumber\\
	&&\quad\quad\quad\quad\quad\quad
	(\Delta_{x_1}-\Delta_{x_1'})\gamma^{(2)}(x_1,x_2;x_1',x_2')
	\nonumber\\
	 & = &- \frac{\mu}{2}\int dx_1 dx_1' \, \delta(x_1-x_1') \, 
	\label{eq-Encons-III-aux-1}\\
	&&\quad\quad\quad\quad\quad\quad
	\Big[ \, (\Delta_{x_1}\gamma^{(2)})(x_1,x_1' ;x_1',x_1' )
	-(\Delta_{x_1'} \gamma^{(2)})(x_1,x_1;x_1',x_1) \, \Big] \,.
	\nonumber
\eeqn
We will show that this term is canceled by the term $(II)$.
\\

\noindent\underline{\em The term $(II)$.}
We have
\eqn
	(II) &=& -\frac\mu2\tr( \, \nabla_{x}\cdot\nabla_{x'}(B_{1,2}\gamma^{(2)}) \, )  
	\\
	&=& -\frac\mu2 \int dxdx'\delta(x-x')
	\nabla_{x}\cdot\nabla_{x'}( \gamma^{(2)}(x_1,x_1;x_1',x_1)-\gamma^{(2)}(x_1,x_1';x_1',x_1'))
	\nonumber
\eeqn
Now we use \eqref{eq-trderA-id-aux-1}, and we obtain
\eqn
	(II) &=& -\frac\mu2\int dx_1dx_1'\delta(x_1-x_1') \label{eq-Encons-II-aux2}
	\\  
	&&\quad\quad\quad\quad\quad\quad
	\Big[ \,
	(\Delta_{x_1'} \gamma^{(2)})(x_1,x_1;x_1',x_1)-
	(\Delta_{x_1}\gamma^{(2)})(x_1,x_1';x_1',x_1')) \, \Big]\,. 
        \nonumber 
	\eeqn
Hence \eqref{eq-Encons-III-aux-1} and \eqref{eq-Encons-II-aux2} imply $(II)=-(III)$.

We conclude that $\partial_t E_1(\Gamma(t))=0$. Therefore, $E_1(\Gamma(t))=E_1(\Gamma(0))$
is a conserved quantity. It represents the average energy per particle. 
For the quintic case $p=4$ (or similarly in more general cases $p\in2\N$), the above arguments
can be adapted straightforwardly.
\qed

\section{Virial identities}
\label{sec-virial}

In this section, we prove the virial identities necessary for the application of a
generalized version of Glassey's 
argument. 
According to our previous discussion, it is sufficient to  consider only
$\gamma^{(1)}(x_1 ; x_{1}^{\prime})$.
 
\subsection{Density}

In what follows, we drop the superscript ``${(k)}$'' from $\gamma^{(k)}$.
It will be clear from the number of variables what the value of $k$ (in this 
part of the discussion, $k=1$ or $k=2$) is in a given expression.

We write
\eqn
	\gamma(x;x') \, = \, \int dv dv' e^{ivx-iv'x'}\widehat\gamma(v;v')
\eeqn
and define
\eqn
	\rho(x) \, := \, \gamma(x;x)
	\, = \, \int dv \, dv' \, e^{i(v-v')x}\widehat\gamma(v;v') \,.
\eeqn
Thus,
\eqn
	\partial_t\rho(x) & = & 
	\int dv \, dv' \, e^{i(v-v')x} \, \partial_t\widehat\gamma(v;v')
        \label{vir-dens-partial}\\
        & = & 
        -\frac{1}{i} \int dv \, dv' e^{i(v-v')x} \, 
        \widehat{ (\Delta_x - \Delta_{x'}) \gamma }(v,v') 
        \nonumber \\   
        &&
        \quad + \, \frac1i\int dvdv'  \, e^{i(v-v')x}\, \widehat{B_{1,2}\gamma}(v;v').
        \nonumber
\eeqn

First we notice that 
\begin{align} 
        & -\frac{1}{i} \int dv \, dv' e^{i(v-v')x} \, 
        \widehat{ (\Delta_x - \Delta_{x'}) \gamma }(v,v') 
        \nonumber \\
        & \quad \quad = 
	\frac1i\int dv \, dv' \, e^{i(v-v')x}(v^2 - (v')^2) \, \widehat\gamma(v;v')
        \nonumber\\
        & \quad \quad = 
	\frac1i\int dv \, dv' \, e^{i(v-v')x}(v+v')(v-v') \, \widehat\gamma(v;v')
        \nonumber\\
        & \quad \quad =  
	\nabla_x \cdot \int dv \, dv' \, e^{i(v-v')x}(v+v') \, \widehat\gamma(v;v') \,.
        \label{vir-dens1}  
\end{align}

On the other hand, 
\eqn 
	B_{1,2}^1\gamma(x;x') & = &
	\int dy \, dy' \, \delta(x-y) \, \delta(x-y') \,
	\nonumber\\
	&&\quad\quad \int dudqdu'dq'
	e^{i(ux+qy-u'x'-q'y')} \, \widehat\gamma(u,q;u',q') 
	\nonumber\\
	 & = &
	\int dudqdu'dq'
	e^{i((u+q-q')x-u'x')} \, \widehat\gamma(u,q;u',q') \label{vir-Bcubic-pos} 
	\,.
\eeqn
Therefore, 
\eqn
	\widehat{B_{1,2}^1\gamma}(v;v')
	& = & \int dxdx' \, e^{-ivx+iv'x'} \, B_{1,2}^1\gamma(x;x') 
	\nonumber\\
	& = & 
	\int dudqdu'dq' \delta(u+q-q'-v) \, \delta(v'-u') \,  \widehat\gamma(u,q;u',q') 
	\nonumber\\
	& = & 
	\int dqdq'  \,  \widehat\gamma(v-q+q',q;v',q') \,. \label{vir-B1cubic-hat}
\eeqn
Likewise, one obtains
\eqn
	\widehat{B_{1,2}^2\gamma}(v;v')
	& = & \int dxdx' \, e^{-ivx+iv'x'} \, B_{1,2}^2\gamma(x;x') 
	\nonumber\\
	& = & 
	\int dudqdu'dq' \delta(v-u) \, \delta(v'-(u'+q'-q)) \,  \widehat\gamma(u,q;u',q') 
	\nonumber\\
	& = & 
	\int dqdq'  \,  \widehat\gamma(v,q;v'+q-q',q') \,. \label{vir-B2cubic-hat} 
\eeqn

Thus,
\eqn
	\lefteqn{
	\frac1i\int dvdv'  \, e^{i(v-v')x}\, \widehat{B_{1,2}\gamma}(v;v')
	}
	\nonumber\\
	& = &
	\frac1i\int dvdv'  \, e^{i(v-v')x}\, (\widehat{B_{1,2}^1\gamma}(v;v') - \widehat{B_{1,2}^2\gamma}(v;v'))
	\nonumber\\
	& = &
	\frac1i\int dvdv' dqdq' \, e^{i(v-v')x}\, \widehat\gamma(v-q+q',q;v',q')
	\nonumber\\
	&&- \, \frac1i\int dvdv' dqdq' \, e^{i(v-v')x}\, \widehat\gamma(v,q;v'+q-q',q') 
        \label{vir-dens2} \\ 
        & = & \,0, \nonumber 
\eeqn
where the last equality is obtained by applying the change 
of variables $v\rightarrow v-q+q'$ and $v'\rightarrow v'-q+q'$ 
in the second term of \eqref{vir-dens2} 
so that the difference $v-v'$ remains unchanged.

Therefore, by combining \eqref{vir-dens-partial}, \eqref{vir-dens1} 
and \eqref{vir-dens2} we conclude 
$$ \partial_t \rho(x) - \nabla_x \cdot P = 0,
$$ 
where 
\eqn
	P \, := \, \int du \, du' \, e^{i(u-u')x}(u+u') \, \widehat\gamma(u;u')
\eeqn
corresponds to the momentum.
\\

\subsection{Morawetz action}

We define
\eqn
	M \, := \, \int dx \, x \cdot P \,.
\eeqn
The time derivative is given by
\eqn
	\partial_t M \, = \, \int dx \, x \cdot \partial_t P \, = \, (I_M) \, + \, (II_M) \,,
\eeqn
where $(I_M)$ is the kinetic, and $(II_M)$ the interaction term.

We have
\eqn
	(I_M) & = & \frac{1}{i} \int dx \, x \cdot\, \int du \, du' \, e^{i(u-u')x}(u+u') \,(u^2 - (u')^2) \, \widehat\gamma(u;u')
	\nonumber\\
	& = & \frac{1}{i} \int dx \, x \cdot\int du \, du' \, e^{i(u-u')x}[(u+u')\otimes(u+u')](u-u') \,  \widehat\gamma(u;u')
	\nonumber\\
	& = & - \, \int du \, du' \,   \widehat\gamma(u;u') \int dx \, x \cdot [(u+u')\otimes(u+u')] (\nabla_x e^{i(u-u')x} )\, 
	\nonumber\\
	& = & \,  \int du \, du' \, \widehat\gamma(u;u') \, \tr[(u+u')\otimes(u+u')] \, \int dx \; e^{i(u-u')x}
	\nonumber\\
	& = & \int du \, du' \, \delta(u-u')\tr[(u+u')\otimes(u+u')] \,  \widehat\gamma(u;u')
	\nonumber\\
	& = & 4 \int du \,u^2  \widehat\gamma(u;u) \,. \label{eq-p-I_M}
\eeqn
which is $8$ times the kinetic energy of one particle.

\subsection{Interaction term}

Next, we study the interaction term 
\eqn
	(II_M) 
	& = & \mu
	\int dx \, x \cdot \frac1i\int dvdv'  \, e^{i(v-v')x} \, (v+v')  \, \widehat{B_{1,2}\gamma}(v;v') \,.
\eeqn

\noindent\underline{\em (A) The cubic case.}


As we have seen in \eqref{vir-Bcubic-pos}, we have
$$
	B_{1,2}^1\gamma(x;x') = 
	\int dudqdu'dq'
	e^{i((u+q-q')x-u'x')} \, \widehat\gamma(u,q;u',q') 
	\,.
$$
Therefore, by \eqref{vir-B1cubic-hat}  
$$
	\widehat{B_{1,2}^1\gamma}(v;v')
	= \int dqdq'  \,  \widehat\gamma(v-q+q',q;v',q') \,.
$$ 
Likewise by \eqref{vir-B2cubic-hat} 
$$
	\widehat{B_{1,2}^2\gamma}(v;v')
	= \int dqdq'  \,  \widehat\gamma(v,q;v'+q-q',q') \,.
$$

Now we determine the term $(II_M)$ in $\int dx \, x \cdot\partial_tP$ that involves the interaction. 
To this end, we first consider
\eqn
	\lefteqn{
	\frac1i\int dvdv'  \, e^{i(v-v')x} \, (v+v')  \, \widehat{B_{1,2}\gamma}(v;v')
	}
	\nonumber\\
	& = &
	\frac1i\int dvdv'  \, e^{i(v-v')x} \, (v+v') \, (\widehat{B_{1,2}^1\gamma}(v;v') - \widehat{B_{1,2}^2\gamma}(v;v'))
	\nonumber\\
	& = &
	\frac1i\int dvdv' dqdq' \, e^{i(v-v')x} \, (v+v') \, \widehat\gamma(v-q+q',q;v',q')
	\nonumber\\
	&&- \, \frac1i\int dvdv' dqdq' \, e^{i(v-v')x} \, (v+v') \, \widehat\gamma(v,q;v'+q-q',q')   \,.
\eeqn
In the last term, we apply the change of variables $v\rightarrow v-q+q'$ and $v'\rightarrow v'-q+q'$,
so that the difference $v-v'$ remains unchanged. 
We obtain that the above equals
\eqn 
	\lefteqn{
	\frac1i\int dvdv' dqdq' \, e^{i(v-v')x} \, (v+v') \, \widehat\gamma(v-q+q',q;v',q')
	}
	\nonumber\\
	&&- \, \frac1i\int dvdv' dqdq' \, e^{i(v-v')x} \, (v+v'-2q+2q') \, \widehat\gamma(v-q+q',q;v',q')
	\nonumber\\
	&=&\frac1i\int dvdv' dqdq' \, e^{i(v-v')x} \,  \widehat\gamma(v-q+q',q;v',q')
	\, \big( \, (v+v') \, - \,  (v+v'-2q+2q') \, \big)
	\nonumber\\
	&=&\frac1i\int dvdv' dqdq' \, e^{i(v-v')x} \, 2(q-q') \, \widehat\gamma(v-q+q',q;v',q') \,.
\eeqn
The contribution of this term to the integral $\int dx \, x\cdot\partial_tP$ is given by
\eqn
	\frac{\mu}{i}\int dx x\cdot \int dvdv' dqdq' \, e^{i(v-v')x} \, 2(q-q') \, \widehat\gamma(v-q+q',q;v',q') \,.
\eeqn
Next, we express everything in position space.

We have that the last line equals
\eqn
	\lefteqn{
	\frac{\mu}{i}\int dx \; x\cdot \int dXdYdX'dY'\int dvdv' dqdq' \, e^{i(v-v')x} \, 2(q-q') \, 
	}
	\nonumber\\
	&&\quad\quad\quad\quad
	e^{i(-(v-q+q')X-qY+v'X'+q'Y')} \gamma(X,Y;X',Y')  
	\nonumber\\
	&=&
	\frac{\mu}{i}\int dx   \int dXdYdX'dY' \, 
	\gamma(X,Y;X',Y')
	\, \int dvdv' dqdq' \,  
	\nonumber\\
	&&\quad\quad\quad\quad
	e^{iv(x-X)-iv'(x-X')} \, 2x\cdot(q-q') \, e^{+iq(X-Y)-q'(X-Y')}\,   \,  
	\nonumber\\ 
	&=&
	-\mu \int dx   \int dXdYdX'dY' \, 
	\gamma(X,Y;X',Y') \int dqdq' \,  
	\nonumber\\
	&&\quad\quad\quad\quad
	\delta(x-X)\delta(x-X') \, 2X\cdot\nabla_X \, e^{+iq(X-Y)-iq'(X-Y')}\,   
	\nonumber\\
	&=&
	-\mu \int dXdYdY' \, 
	\gamma(X,Y;X ,Y')  
	\nonumber\\
	&&\quad\quad\quad\quad  
	\, 2X\cdot\nabla_X \, \delta(X-Y) \, \delta(Y-Y') \, 
	\label{eq-nonlinterm-aux-1}\\
	&=&
	-\mu \int dXdY  \, 
	\gamma(X,Y;X ,Y)   
	\, 2X\cdot\nabla_X \, \delta(X-Y)  
	\nonumber\\
	&=&
	\mu \int dXdY  \, \delta(X-Y) \, (2d+ 2X\cdot\nabla_X \,  )
	\gamma(X,Y;X ,Y)   
	\label{eq-nonlinterm-aux-2}
\eeqn
where we have written $\delta(X-Y)\delta(X-Y')=\delta(X-Y)\delta(Y-Y')$ to get (\ref{eq-nonlinterm-aux-1}).

Now we note that
\eqn
	\lefteqn{
	\int dX \, 
	X\cdot\nabla_X\gamma(X,X;X ,X)   
	}
	\label{vir-int-cubic1}\\
	&=&
	\int dXdY  \, \delta(X-Y) \, ( \,  X\cdot\nabla_X \,  + \, Y\cdot\nabla_Y \, )
	\gamma(X,Y;X ,Y)   
	\nonumber\\
	&=&
	\int dXdY  \, \delta(X-Y) \, ( \,  X\cdot\nabla_X \gamma(X,Y;X ,Y) \,  + \, Y\cdot\nabla_Y \gamma(Y,X;Y,X) \, )
	\nonumber\\
	&=&
	\int dXdY  \, \delta(X-Y) \, ( \,  2X\cdot\nabla_X  \gamma(X,Y;X ,Y)  \, ) \label{eq-cubic-div}
\eeqn
where we used the symmetry $\gamma(X,Y;X,Y)=\gamma(Y,X;Y,X)$, and renamed the variables in the last term.
Clearly, \eqref{vir-int-cubic1} equals
\eqn  \label{eq-cubic-dim}
	-d\int dX \,  \gamma(X,X;X ,X)  
\eeqn
from integrating by parts.

Therefore, combining \eqref{eq-nonlinterm-aux-2},  \eqref{eq-cubic-div} and  \eqref{eq-cubic-dim}
\eqn
	(II_M)
	& = & \mu \int dXdY  \, \delta(X-Y) \, (2d+ 2X\cdot\nabla_X \,  )
	\gamma(X,Y;X ,Y)
	\nonumber\\
	& = & \mu \int dXdY  \, \delta(X-Y) \, (2d - d \,  )
	\gamma(X,Y;X ,Y)
	\nonumber\\ 
	& = & \mu d\int dX \, 
	\gamma(X,X;X ,X) \,.
\eeqn 
This is the desired result for the cubic case.
\\

\noindent\underline{\em (B) The quintic case.}

Now we give a sketch of the calculations related to the interaction term in 
the quintic ($p=4$) case. Again it suffices to consider $k=1$. 

Since in the case when $p=4$ we have   
\eqn
	\lefteqn{
	B_{1;2,3}^1\gamma(x;x') \, = \,
	\int dy \, dy' \, dz \, dz' \delta(x-y) \, \delta(x-y') \, 
                                    \delta(x-z) \, \delta(x-z') \,
   	}     
	\nonumber\\
	&&\quad\quad \int dudqdrdu'dq'dr'
	e^{i(ux+qy+rz-u'x'-q'y'-r'z')} \, \widehat\gamma(u,q,r;u',q',r') 
	\nonumber\\
	 & = &
	\int dudqdrdu'dq'dr'
	e^{i((u+q+r-q'-r')x-u'x')} \, \widehat\gamma(u,q,r;u',q',r') 
	\,,
\eeqn
by taking the Fourier transform we obtain 
\eqn
	\widehat{B_{1;2,3}^1\gamma}(v;v')
	& = & \int dxdx' \, e^{-ivx+iv'x'} \, B_{1;2,3}^1\gamma(x;x') 
	\nonumber\\
	& = & 
	\int dudqdrdu'dq'dr' \delta(u+q+r-q'-r'-v) \,
	\nonumber \\
        &&\quad\quad\quad\quad\quad\quad\quad\quad\quad\quad
        \delta(v'-u') \,  
        \widehat\gamma(u,q,r;u',q',r') 
	\nonumber\\
	& = & 
	\int dqdrdq'dr'  \,  \widehat\gamma(v-q-r+q'+r',q,r;v',q',r') \,.
\eeqn

As in the cubic case, to determine the term $(II_M)$ in $\int dx \, x \cdot\partial_tP$, 
we first observe that 
\eqn
	\lefteqn{
	\frac{1}{i} \int dvdv'  \, e^{i(v-v')x} \, (v+v')  \, \widehat{B_{1;2,3}\gamma}(v;v')
	}
	\nonumber\\
	& = &
	\frac1i\int dvdv'  \, e^{i(v-v')x} \, (v+v') \, (\widehat{B_{1;2,3}^1\gamma}(v;v') - \widehat{B_{1;2,3}^2\gamma}(v;v'))
	\nonumber\\
	& = &
	\frac1i\int dvdv' dqdq'drdr' \, e^{i(v-v')x} \, (v+v') \, 
        \widehat\gamma(v-q-r+q'+r',q,r;v',q',r')
	\nonumber\\
	&&- \, \frac1i\int dvdv' dqdq'drdr' \, e^{i(v-v')x} \, (v+v') \, 
        \widehat\gamma(v,q,r;v'+q+r-q'-r',q',r')   \,,
         \nonumber
 \eeqn
which after performing the change of variables 
$v\rightarrow v-q-r+q'+r'$ and $v'\rightarrow v'-q-r+q'+r'$ in the last term,
becomes 
$$
	\frac1i\int dvdv' dqdq'drdr' \, 
        e^{i(v-v')x} \, 2(q+r-q'-r') \, \widehat\gamma(v-q-r+q'+r',q,r;v',q',r')\;.
$$
Hence the contribution of this term to the integral $\int dx \, x\cdot\partial_tP$ is given by
$$
	\frac{\mu}{i}\int dx x\cdot \int dvdv' dqdq'drdr' \, 
        e^{i(v-v')x} \, 2(q+r-q'-r') \, \widehat\gamma(v-q-r+q'+r',q,r;v',q',r') \,,
$$
which we express in the position space as follows 
\eqn
	\lefteqn{
	\frac{\mu}{i}\int dx \; x\cdot \int dXdYdZdX'dY'dZ'
        \int dvdv' dqdq' drdr'\, e^{i(v-v')x} \, 2(q+r-q'-r') \, 
	}
	\nonumber\\
	&&\quad\quad\quad\quad
	e^{i(-(v-q-r+q'+r')X-qY-rZ+v'X'+q'Y'+r'Z')} \gamma(X,Y,Z;X',Y',Z')  
	\nonumber\\
	&=&
	\frac{\mu}{i}\int dx   \int dXdYdZdX'dY'dZ' \, 
	\gamma(X,Y,Z;X',Y',Z')
	\, \int dvdv' dqdq' drdr'\,  
	\nonumber\\
	&&\quad\quad\quad\quad
	e^{iv(x-X)-iv'(x-X')} \, 2x\cdot(q+r-q'-r') \, 
	\nonumber \\
        &&\quad\quad\quad\quad\quad\quad\quad\quad
        e^{+iq(X-Y)-iq'(X-Y')}\, e^{+ir(X-Z)-ir'(X-Z')}\,    
	\nonumber\\ 
	&=&
	- \mu \int dx   \int dXdYdZdX'dY'dZ' \, 
	\gamma(X,Y,Z;X',Y',Z') \int dqdq'drdr' \,  
	\nonumber\\
	&&\quad\quad\quad\quad
	\delta(x-X)\delta(x-X') \, 2X\cdot\nabla_X \, e^{+iq(X-Y)+ir(X-Z)-iq'(X-Y')-ir'(X-Z')}\,   
	\nonumber\\
	&=&
	-\mu \int dXdYdZdY'dZ' \, 
	\gamma(X,Y,Z;X,Y',Z')  
	\nonumber\\
	&&\quad\quad\quad\quad  
	\, 2X\cdot\nabla_X \, \delta(X-Y) \, \delta(Y-Y') \; \delta(X-Z)\, \delta(Z-Z') 
	\label{quintic-eq-nonlinterm-aux-1}\\
	&=&
	-\mu \int dXdYdZ  \, 
	\gamma(X,Y,Z;X,Y,Z)   
	\, 2X\cdot\nabla_X \, \delta(X-Y)\delta(X-Z)  
	\nonumber\\
	&=&
	\mu \int dXdYdZ  \, \delta(X-Y)\delta(X-Z) \, (2d+ 2X\cdot\nabla_X \,  )
	\gamma(X,Y;X ,Y)   
	\label{quintic-eq-nonlinterm-aux-2}
\eeqn
where we have written 
$$\delta(X-Y)\delta(X-Y')\delta(X-Z)\delta(X-Z')
=\delta(X-Y)\delta(Y-Y')\delta(X-Z)\delta(Z-Z')$$ 
to get (\ref{quintic-eq-nonlinterm-aux-1}).

On the other hand, using symmetry of $\gamma(X,Y,Z;X,Y,Z)$ we obtain 
\eqn
	\lefteqn{
	\int dX \, X\cdot\nabla_X\gamma(X,X,X;X,X,X)
        }	
        \nonumber\\
	&=&
	\int dXdYdZ  \, \delta(X-Y)\delta(X-Z) \, \nonumber \\
        &&\quad\quad\quad\quad 
        ( \,  X\cdot\nabla_X \,  + \, Y\cdot\nabla_Y \,+ \, Z\cdot\nabla_Z \, )
	\gamma(X,Y,Z;X,Y,Z)   
	\nonumber\\
	&=&
	\int dXdYdZ  \, \delta(X-Y)\delta(X-Z) \,\nonumber \\
        &&\quad\quad\quad\quad 
        ( \,  X\cdot\nabla_X \gamma(X,Y,Z;X,Y,Z) \,  
         + \, Y\cdot\nabla_Y \gamma(Y,X,Z;Y,X,Z) \,
         \nonumber\\
         &&\quad\quad\quad\quad\quad\quad\quad\quad\quad\quad\quad\quad\quad\quad\quad\quad\quad\quad 
         + \, Z\cdot\nabla_Z \gamma(Z,Y,X;Z,Y,X) \,)
	\nonumber\\
	&=&
	\int dXdYdZ  \, \delta(X-Y)\delta(X-Z) \, ( \,  3X\cdot\nabla_X  \gamma(X,Y,Z;X,Y,Z)  \,. ) \label{renaming} 
\eeqn \
However, by the integration by parts, 
\eqn \label{quintic-lhs-beforesym}
        \int dX \, X\cdot\nabla_X\gamma(X,X,X;X,X,X) = 
	-d\int dX \,  \gamma(X,X,X;X,X,X),  
\eeqn
so by combining \eqref{renaming} and \eqref{quintic-lhs-beforesym} we obtain 
\eqn \label{quintic-dimdepen}
        \int dXdYdZ  \, \delta(X-Y)\delta(X-Z) \, ( \, X\cdot\nabla_X  \gamma(X,Y,Z;X,Y,Z) 
        \nonumber \\
        \quad \quad = -\frac{d}{3} \int dX \,  \gamma(X,X,X;X,X,X).
\eeqn
Therefore
\eqn
	(II_M)
	& = & \mu \int dXdYdZ  \, \delta(X-Y)\delta(X-Z) \, (2d+ 2X\cdot\nabla_X \,  )
	\gamma(X,Y,Z;X,Y,Z)
	\nonumber\\
	& = & \mu \int dXdYdZ  \, \delta(X-Y)\delta(X-Z) \, (2d - 2d/3) \,  )
	\gamma(X,Y;X ,Y)
	\nonumber\\ 
	& = & \mu \frac{4d}{3} \int dX \, 
	\gamma(X,X,X;X,X,X) \,.
\eeqn 
This is the desired result for the quintic case.
\\

\noindent\underline{\em (C) The general case $p \in 2{\mathbb N}$ }

The above calculation on the interaction term can be reproduced for a general even number $p$
and it that case we obtain:  
\eqn  
          (II_M) 
	& = & \mu \int dX_1 dX_2 ... dX_{1+\frac p2} \, \delta(X_1 - X_2) ... \delta(X_1 - X_{1+\frac p2}) 
	\nonumber \\
	&&\quad \quad \ (2d+ 2X_1\cdot\nabla_{X_1} \,  )
	\gamma(X_1,...,X_{1+\frac p2};X_1,...,X_{1+\frac p2})
	\nonumber\\
	& = & \mu \int dX_1 dX_2 ... dX_{1+\frac p2} \, \delta(X_1 - X_2) ... \delta(X_1 - X_{1+\frac p2}) 
	\nonumber \\
          &&\quad \quad \ (2d - 2\frac{d}{1+\frac p2} \,  )
	\gamma(X_1,...,X_{1+\frac p2};X_1,...,X_{1+\frac p2})
	\nonumber\\
	& = & \mu \; \frac{2dp}{p+2} \int dX \, 
         \gamma( \underbrace{X,...,X}_{1+\frac p2}; \underbrace{X,...,X}_{1+\frac p2})\,. \label{eq-p-II_M}
\eeqn

Now we combine \eqref{eq-p-I_M}  and \eqref{eq-p-II_M} to conclude that: 
\eqn\label{eq-virial-final-1}
	\lefteqn{
	\partial_t^2\int dx \, x^2\gamma(x,x) 
	}
	\nonumber \\
	&=& 2 \int dx \, x\cdot\partial_tP
	\nonumber\\
	& = &
	8\int du \, u^2 \, \widehat\gamma(u;u) \, 
	+ \, \mu \; \frac{4dp}{p+2} \int dX \, 
         \gamma( \underbrace{X,...,X}_{1+\frac p2}; \underbrace{X,...,X}_{1+\frac p2})\,.
\eeqn

\section{Glassey's argument and blowup in finite time} 
\label{sec-Glassey}

Now we are prepared to prove blowup in finite time for negative energy initial
conditions, by generalizing Glassey's argument familiar from NLS and related
nonlinear PDE's, to the GP hierarchy.

The quantity that will be relevant in 
reproducing   Glassey's argument is given by 
\eqn\label{eq-Vk-def-1}
	V_k( \, \Gamma(t) \, ) & := & \tr( \, \sum_{j=1}^k \, x_j^2 \gamma^{(k)}(t) \, )  \,.
\eeqn
Similarly as in our discussion of the conserved energy, we observe that 
\eqn
	V_k( \, \Gamma(t) \, ) & =&  \tr( \, \sum_{j=1}^k \, x_j^2 \gamma^{(k)}(t) \,)  
    \nonumber\\
                               & =& k \; \tr ( \, x_1^2 \gamma^{(1)}(t) \,)
    \nonumber\\
                               & =& k \; V_1( \, \Gamma(t) \, )\,.
	\label{eq-Vk1} 
\eeqn
Again, this follows from the fact that $\gamma^{(k)}$ is symmetric in its variables,
and from the admissibility of $\gamma^{(k)}(t)$ for all $k\in\N$,

Next, we relate $\partial_{t}^2 V_1(t)$ to the conserved
energy per particle. First, let us denote by $E_1^K(t)$ the kinetic part of the energy 
$E_1(t)$ and by $E_1^P(t)$ the potential part of the energy $E_1(t)$ i.e. 
\eqn
            E_1^K(\, \Gamma(t) \,) & =&  \frac12\tr( \, (-\Delta_{x}) \gamma^{(1)}(t)  \, ),  
            \nonumber \\	
            E_1^P(\, \Gamma(t) \,) & =&  \frac{\mu}{p+2} 
            \tr( \, B_{1;2,\dots,1+\frac p2}^1 \, \gamma^{(1+\frac p2)}(t) \, ).  
\eeqn
From (\ref{eq-virial-final-1}), we can relate $\partial_{t}^2 V_1(t)$ to the conserved
energy per particle as follows
\eqn
	\partial_{t}^2 V_1(t) 
	& = &
	8\int du \, u^2 \, \widehat\gamma(u;u) \, 
	+ \, \mu \; \frac{4dp}{p+2} \int dX \, 
         \gamma( \underbrace{X,...,X}_{1+\frac p2}; \underbrace{X,...,X}_{1+\frac p2}) 
         \nonumber\\
         & = &
         16 E_1^K(\, \Gamma(t) \,)  + 4dp \, E_1^P(\, \Gamma(t) \,) 
        \nonumber\\
        & = & 
        16 E_1(\, \Gamma(t) \,) + 4d \, (p-\frac 4d) \, E_1^P(\, \Gamma(t) \,) 
        \nonumber \\        
        & = & 
        16 E_1(\, \Gamma(0) \,)  + 4d \mu \, \frac{p - p_{L^2}}{p+2}  
         \int dX \, 
         \gamma( \underbrace{X,...,X}_{1+\frac p2}; \underbrace{X,...,X}_{1+\frac p2}),
         \label{eq-Glassey-p-final}
 \eeqn 
where we used the fact that $E_1( \, \Gamma(t) \, )$ is conserved.

Now we conclude that for the focusing ($\mu = -1$)  GP hierarchy which is either at the 
$L^2$-critical level ($p = p_{L^2}$) or at the $L^2$-supercritical ($p > p_{L^2}$) level, 
\eqn\label{eq-negcurv-V-1}
	\partial_{t}^2 V_1(t)  \, \leq \, 16 E_1(\, \Gamma(0) \,).
\eeqn 
However, the function 
$V_1(t)$ is nonnegative, so we conclude that if $E_1( \, \Gamma(0) \, )<0$, the solution 
blows up in finite time.

To be precise, we infer from (\ref{eq-negcurv-V-1}) that there exists a finite
time $T^*$ such that
$V_1(t)\searrow0$ as $t\nearrow T^*$. Accordingly,
\eqn
	1 & = & \tr ( \, \gamma^{(1)}(t) \, )
	\nonumber\\
	& \leq & ( \, \tr ( \,  x^2\gamma^{(1)}(t) \, ) \, )^{1/2} 
	(\tr ( \, \frac{1}{x^2}\gamma^{(1)}(t) \, ) \, )^{1/2}
	\nonumber\\
	& \leq &C \, (\tr ( \, x^2\gamma^{(1)}(t) \, ) \, )^{1/2} 
	(\tr( \, -\Delta\gamma^{(1)}(t) \, ) )^{1/2}
\eeqn
where we have first used the Cauchy-Schwarz, and subsequently the Hardy inequality.
Thus, $\tr (-\Delta\gamma^{(1)}(t) )\geq (V_1(t))^{-1} \nearrow\infty$ as $t\nearrow T^*$.

One can easily verify from (\ref{def-Sobnorms-1}) that
\eqn
	\| \, \gamma^{(k)}(t) \, \|_{\cH^1_k} & \geq &
	\sum_{j=1}^k\tr(-\Delta_{x_j}\gamma^{(k)}(t) )
	\, + \, \sum_{j=1}^k\tr(-\Delta_{x_j'}\gamma^{(k)}(t) )
	\nonumber\\
	&=&2k \tr (-\Delta\gamma^{(1)}(t) ) \; \nearrow\infty
\eeqn
as $t\nearrow T^*$.
Accordingly, $\Av_{H^1}(\Gamma(t)) \nearrow\infty$ as $t\nearrow T^*$, which 
establishes blowup in finite time.

\subsection*{Acknowledgments} 
T.~C. thanks I. Rodnianski for inspiring discussions.
The work of T.~C. was supported by NSF grant DMS-0704031.
The work of N.~P. was supported by NSF grant number DMS 0758247 
and an Alfred P. Sloan Research Fellowship. The work of N.~T.  was supported by NSF grant DMS-0901222.

\end{document}